\documentclass[10pt]{extarticle}
\usepackage{spconf}
\usepackage{cite}
\usepackage{amsmath,amssymb,amsfonts}
\usepackage{algorithmic}
\usepackage{graphicx}
\usepackage{textcomp}
\usepackage{xcolor}
\def\BibTeX{{\rm B\kern-.05em{\sc i\kern-.025em b}\kern-.08em
    T\kern-.1667em\lower.7ex\hbox{E}\kern-.125emX}}

\usepackage{algorithm}
\usepackage{array}
\usepackage{stfloats}
\usepackage{url}
\usepackage{verbatim}
\usepackage{amsmath,amsthm}   
\usepackage{amssymb}
\usepackage{nomencl}
\usepackage{multirow}

\usepackage{bm}
\usepackage{lipsum}
\usepackage{makeidx}
\usepackage{enumerate}
\usepackage{color}

\usepackage[font=small,labelfont=bf]{caption}
\usepackage[font=small,labelfont=bf]{subcaption}

\usepackage{tikz,pgfplots}
\usetikzlibrary{calc}
\usetikzlibrary{intersections}
\usetikzlibrary{arrows,shapes}
\usetikzlibrary{positioning}

\newtheorem{corollary}{Corollary}

\newtheorem{lemma}{Lemma}

\theoremstyle{definition}

\allowdisplaybreaks

\begin{document}
\title{
Maximization of minimum rate in MIMO OFDM RIS-assisted  Broadcast Channels
}
%\vspace{-.7cm}
\name{Mohammad Soleymani$^1$, Ignacio Santamaria$^2$, Aydin Sezgin$^3$, and Eduard Jorswieck$^4$%\vspace{-.2cm}
\thanks{
The work of Ignacio Santamaria was funded by MCIN/ AEI
/10.13039/501100011033 under Grant PID2022-137099NB-C43 (MADDIE). 
The work of Eduard Jorswieck was supported in part by the Federal Ministry of Education and Research (BMBF, Germany) as part of the 6G Research and Innovation Cluster 6G-RIC under Grant 16KISK031.
This work of Aydin Sezgin is funded by the German Federal Ministry of Education and
Research (BMBF) in the course of the 6GEM Research Hub under Grant
16KISK037.
}}
\address{$^1$Signal and System Theory Group, Universit\"at Paderborn, Germany\\ $^2$Department of Communications Engineering, University of Cantabria\\  
$^3$Digital Communication Systems, Ruhr University Bochum, Germany\\
$^4$ Institute for Communications Technology, Technische Universit\"at Braunschweig,  Germany\\
Emails: %\small
{\protect\url{mohammad.soleymani@sst.upb.de,}} %\small
{\protect\url{i.santamaria@unican.es}}, 
\\
%\small
{\protect\url{aydin.sezgin@rub.de}},
%\small
{\protect\url{e.jorswieck@tu-bs.de}}
%\vspace{-.6cm}
}
\maketitle
\begin{abstract}
Reconfigurable intelligent surface (RIS) is a promising technology to enhance the spectral efficiency of wireless communication systems. By optimizing the RIS elements, the performance of the overall system can be improved. Yet, in contrast to single-carrier systems, in multi-carrier systems, it is not possible to independently optimize RIS elements  at each sub-carrier, which may reduce the benefits of RIS in multi-user orthogonal frequency division multiplexing (OFDM) systems. To this end, we investigate the effectiveness of RIS in multiple-input, multiple-output (MIMO) OFDM broadcast channels (BC). We formulate and solve a joint precoding and RIS optimization problem. We show that RIS can significantly improve the system performance even when the number of RIS elements per sub-band is very low.  
\end{abstract} 
%{IEEEkeywords}%
%\vspace{-.4cm}

\section{Introduction}
%\vspace{-.2cm}
Intelligent metasurfaces can help us to realize smart radio environments in which the channels are not determined by nature only \cite{wu2019towards}. Instead, the channels can be optimized, which provides additional degrees of freedom in the system design, and can yield a huge performance improvement \cite{wu2021intelligent, di2020smart}. Among different technologies for intelligent metasurfaces is the passive reflective reconfigurable intelligent surface (RIS). It has been shown that RIS can be a powerful technology to improve the spectral and energy efficiency of various single-carrier systems, including point-to-point communication systems \cite{zappone2020optimal}, broadcast channels (BCs) \cite{huang2019reconfigurable, wu2019intelligent, soleymani2022rate, soleymani2023rate, pan2020multicell, soleymani2022improper, soleymani2022noma, soleymani2023energy}, cognitive radio systems \cite{zhang2020intelligent}, $K$-user interference channels \cite{jiang2022interference, santamaria2023icassp}, cell-free multiple-input, multiple-output (MIMO) systems \cite{weinberger2023ris}, cloud radio access networks  \cite{esmaeili2022fairness, weinberger2022synergistic} and ultra-reliable low-latency communication systems \cite{soleymani2023spectral, soleymani2023optimization}. 
 
Yet, single-carrier systems require a demanding equalization process in case the channel is frequency selective. As an alternative, we may have to employ techniques such as orthogonal frequency division multiplexing (OFDM) to use frequency bands more efficiently. Unfortunately, when OFDM is employed, RIS elements cannot be independently optimized  at each subband \cite{li2021intelligent}. In a more realistic scenario, the RIS elements remain approximately constant in all frequency subbands, which is highly suboptimal when the number of subbands grows. Thus, one might expect that the benefits of RIS may disappear when the bandwidth is large. 

The performance of RIS in OFDM systems has been studied in \cite{zheng2019intelligent,  yang2020risofdm, kompostiotis2023received, li2021intelligent}. The authors in \cite{zheng2019intelligent} considered a point-to-point single-input single-output (SISO) RIS-assisted OFDM system and proposed schemes to estimate channels as well as to optimize RIS elements. In \cite{yang2020risofdm}, the authors developed a practical transmission protocol for point-to-point SISO RIS-assisted OFDM systems. In \cite{kompostiotis2023received}, the authors studied a multi-user SISO RIS-assisted OFDM BC and optimized the RIS elements to maximize the total power received by users for a given transmit power. Finally, the authors in \cite{li2021intelligent} showed that RIS can increase the sum rate of a multi-user multiple-input single-output (MISO) RIS-assisted OFDM BC. Even though these studies provide a valuable insight on the performance of RIS in OFDM systems, they did not consider multi-user MIMO OFDM systems, which is the focus of this paper.

In this paper, we propose a resource allocation scheme to maximize the minimum rate of users in a multiple-input multiple-output (MIMO) RIS-assisted OFDM BC. In contrast to state of the art, we need a jointly optimization of the transmit covariance matrices for each subband at the base station (BS) and RIS elements. We show that RIS can significantly increase the minimum rate of users even with a relatively low number of RIS components per subband. In more details, our numerical results show that optimizing the RIS elements can provide a significant gain in OFDM systems even when the RIS elements cannot be independently optimized at each subband. These results show that RIS can be a promising technology to enhance the performance of OFDM systems. 

\section{System model}
We consider a BC in which there is one BS with $N_{\mathsf{BS}}$ antennas, serving $K$ users with $N_{\mathsf{U}}$ antennas each. Additionally, there is a RIS with $N_{\mathsf{RIS}}$ elements that assists the BS. The BS transmits a linear superposition of multicarrier OFDM signals with $N_i$ subcarriers. In the following, we represent the RIS model as well as the signal model and the rate expressions. Finally, we state the considered optimization problem.

\subsection{RIS model}
We consider the channels in frequency domain. To this end, we employ the channel model in \cite[Eq. (12)]{li2021intelligent}. Thus, the channel between the BS and user $k$ at subband $i$ is 
\begin{equation}\label{ch-equ}
\mathbf{H}_{ki}\left(\bm{\Theta}\right)=
\underbrace{%\sum_{m=1}^M
\mathbf{G}_{ki}\bm{\Theta}\mathbf{G}_{i}}_{\text{Link through RIS}}
+
\underbrace{\mathbf{F}_{ki}}_{\text{Direct link}}\in\mathbb{C}^{N_{\mathsf{U}}\times N_{\mathsf{BS}}}, 
\end{equation}
where $\mathbf{G}_{ki}\in\mathbb{C}^{N_{\mathsf{U}}\times N_{\mathsf{RIS}}}$ is the channel between user $k$ and the RIS in subband $i$, 
$\mathbf{G}_{i}\in\mathbb{C}^{N_{\mathsf{RIS}}\times N_{\mathsf{BS}}}$ is the channel between the RIS and the BS in subband $i$, $\mathbf{F}_{ki}\in\mathbb{C}^{N_{\mathsf{U}}\times N_{\mathsf{BS}}}$ is the direct link between user $k$ and the BS in subband $i$, and $\bm{\Theta}$ is a diagonal matrix containing the RIS coefficients as 
\begin{equation}
\bm{\Theta}=\text{diag}\left(\theta_{1}, \theta_{2},\cdots,\theta_{{N_{\mathsf{RIS}}}}\right).
\end{equation}
We consider a unit modulus constraint for the RIS coefficients, which results in the following feasibility set for $\bm{\Theta}$:
\begin{equation}\label{t-3}
\mathcal{T}=\left\{\theta_{m}:|\theta_{m}|= 1 \,\,\,\forall m\right\}.
\end{equation}
Note that the unit modulus constraint is based on the analytical assumption of operating in a passive and lossless mode, which may not be practical. In \cite[Sec. II.B]{wu2021intelligent}, more practical feasibility sets are presented for RIS elements, which can be studied in a future work.
Additionally, note that according to the model in \cite{li2021intelligent}, the reflecting coefficients cannot be independently optimized at each subband. Thus, we assume that the coefficients remain constant at all the frequency subbands. 

Note that as can be verified through \eqref{ch-equ}, the channels are linear functions of $\bm{\Theta}$. However, for notational simplicity, we do not state this explicitly hereafter.

\subsection{Signal model and rate expression}
The signal transmitted by the BS at subband $i$ is
\begin{equation*}
\mathbf{x}_{i}=\sum_{k=1}^K\mathbf{x}_{ki}\in\mathbb{C}^{N_{\mathsf{BS}}\times 1},
\end{equation*}
where $\mathbf{x}_{ki}\in\mathbb{C}^{N_{\mathsf{BS}}\times 1}$ is the transmit  signal intended for user $k$ at subband $i$. The signals, $\mathbf{x}_{ki}\sim\mathcal{CN}({\bf 0},{\bf P}_{ki})$, are zero-mean, independent and Gaussian random vectors, where ${\bf P}_{ki}=\mathbb{E}\{\mathbf{x}_{ki}\mathbf{x}_{ki}^H\}$ is the covariance matrix of $\mathbf{x}_{ki}$. The received signal at user $k$ in subband $i$ is
\begin{align*}
%\underline
{\mathbf{y}}_{ki}&=
%\underline
{\mathbf{H}}_{ki}
\sum_{j=1}^K%\underline
{\mathbf{x}}_{ji}
+
%\underline
{\mathbf{n}}_{ki},
\end{align*}
where ${\mathbf{n}}_{ki}$ is additive zero-mean Gaussian noise at user $k$ in subband $i$, and ${\mathbf{H}}_{ki}$ is the equivalent channel at subband $i$, given by \eqref{ch-equ}. 

The achievable rate at user $k$ is the summation of the  rates at each subband as
\begin{equation}\label{rate-eq}
r_{k}=\sum_{i=1}^{N_i}r_{ki},
\end{equation}
 where $r_{ki}$ is the rate of decoding $\mathbf{x}_{ki}$ at user $k$.  Treating interference as noise (TIN) at the receiver of user $k$, $r_{ki}$ can be obtained as 
\begin{align}\label{eq-28}
r_{ki}\!
=
\!\log_2\!\left|
{\bf I}\!+\!
{\bf D}_{ki}^{-1}
 {\bf S}_{ki}
\right|\!
=%\!\!
%r_{lk_1,i}\!-r_{lk_2,i},
%\\ &
\underbrace{
%\frac{1}{2}
\!\log_2
\!
\left|{\bf D}_{ki}\!
\!+\!
{\bf S}_{ki}
\right|
}_{r_{ki_1}
}
\!-\!
\underbrace{
%\frac{1}{2}\!
\log_2\left|{\bf D}_{ki}
\right|}
_{r_{ki_2}
},
%\label{eq-29}
\end{align}
where 
%\begin{multline*}
$\mathbf{D}_{ki}
=
\sigma^2{\bf I}+
\sum_{j= 1,j\neq k}^{K}\!\!
%\underline
{\mathbf{H}}_{ki}
\mathbf{P}_{ji}
%\underline
{\mathbf{H}}_{ki}^H$ is the covariance matrix of the noise plus interference,
%}_{\text{Intracell interference}}
%+
%\underbrace{
%\underline{\mathbf{C}}_{i}}_{\text{Noise}},
%\end{multline*} 
and ${\bf S}_{ki}=%\underline
{{\bf H}}_{ki}
{\bf P}_{ki}%\underline
{{\bf H}}_{ki}^H$ is the covariance matrix of the useful signal at user $k$.

\subsection{Problem statement}
In this paper, we aim at maximizing the minimum rate of users, which yields the following optimization problem
%\begin{subequations}\label{ar-opt}
\begin{align}\label{ar-opt}
 \underset{\{\mathbf{P}\}\in\mathcal{P},\bm{\Theta}\in\mathcal{T},r
 }{\max}  & 
  r &
  \text{s.t.}  \,\, \,&  r_k\left(\left\{\mathbf{P}\right\}\!,\!\{\bm{\Theta}\}\right)\geq r,\,\forall k,
  %\\ & \text{Tr}(\mathbf{P}_k)\leq P_k,&k=1,2,
%\\
%\label{4-c}
% &&&
%r_{lk}\geq r^{th},\,\,\,\,\forall l,k,
 \end{align}
%\end{subequations}
where $\bm{\Theta}$ and $\left\{\mathbf{P}\right\}=\left\{\mathbf{P}_{ki}\right\}_{\forall k,i}$ are the optimization variables, $\mathcal{T}$ is %the feasibility set of RIS elements, 
given by \eqref{t-3}, and $\mathcal{P}$ is the feasibility set of the transmit covariance matrices, given by 
\begin{align}
\mathcal{P}&\!=\left\{\{\mathbf{P}\}:\sum_{\forall k,i}\text{Tr}\left(\mathbf{P}_{ki}\right)\leq P, \mathbf{P}_{ki}\succcurlyeq\mathbf{0}, \forall k,i\right\},
\end{align}
where $P$ is the power budget of the BS. 
The minimum rate of users is also referred to as the fairness rate since it is highly probable that all users get the same rate when the minimum rate of users is maximized \cite{soleymani2022noma}.

\section{Proposed optimization framework}
The optimization problem in \eqref{ar-opt} is non-convex. To solve it, we employ majorization minimization (MM) and alternating optimization (AO). To this end, we first fix the RIS coefficients to $\bm{\Theta}^{(t-1)}$, and solve \eqref{ar-opt} to obtain $\left\{\mathbf{P}^{(t)}\right\}$. We then fix the transmit covariance matrices to $\left\{\mathbf{P}^{(t)}\right\}$ and update $\bm{\Theta}$ by solving \eqref{ar-opt}. 
Even in this case, the corresponding optimization problems are still non-convex and complicated to be solved. In the following, we describe our proposed scheme to update the transmit covariance matrices as well as the RIS elements.  

\subsection{Optimizing transmit covariance matrices} 
Now, for given RIS components, $\bm{\Theta}^{(t-1)}$, \eqref{ar-opt} is equivalent to 
%\begin{subequations}\label{ar-opt}
\begin{align}\label{ar-opt-p}
 \underset{\{\mathbf{P}\}\in\mathcal{P},r
 }{\max}  & 
  r &
  \text{s.t.}  \,\, \,&  r_k\left(\left\{\mathbf{P}\right\}\!,\!\bm{\Theta}^{(t-1)}\right)\geq r,\,\forall k,
 \end{align}
%\end{subequations}
which is non-convex since $r_k\left(\left\{\mathbf{P}\right\}\!,\!\bm{\Theta}^{(t-1)}\right)$ for all $k$ is not concave in $\left\{\mathbf{P}\right\}$. 
We employ an MM-based scheme to solve \eqref{ar-opt-p}. That is, we first obtain suitable concave lower bounds for $r_k\left(\left\{\mathbf{P}\right\}\!,\!\bm{\Theta}^{(t-1)}\right)$, which results in a convex surrogate optimization problem. To this end, we employ the lower bound in \cite[Lemma 3]{soleymani2022rate}. For the ease of readers, we restate the lower bound in the following lemma. 
\begin{lemma} %{\bf rewrite it to include the complex domain}
\label{lem-1} 
Consider  positive semi-definite matrices
$\mathbf{A}$ and $\mathbf{B}_j$ for $j=1,\cdots,J$, and arbitrary matrices $\mathbf{C}_j$ for $j=1,\cdots,J$.  
Then, the following inequality holds for all feasible $\{\mathbf{B}_j\}_{\forall j}$ 
\begin{multline*}%\label{eq-lem-1}
\ln\left|\mathbf{A}+\sum_{j=1}^J\mathbf{C}_j\mathbf{B}_j\mathbf{C}^H_j\right| \leq \ln\left|\mathbf{A}+\sum_{j=1}^J\mathbf{C}_j\mathbf{B}_j^{(t)}\mathbf{C}^H_j\right|
\\
 +\!\sum_{j=1}^J\!\mathfrak{R}\!\!\left\{\!\! \text{\em{Tr}}\!\!\left(\!\!\mathbf{C}^H_j\!\!\left(\!\mathbf{A}+\!\!\sum_{j=1}^J\mathbf{C}_j\mathbf{B}^{(t)}_j\mathbf{C}^H_j\!\right)^{-1}\!\!\!\mathbf{C}_j(\mathbf{B}_j-\mathbf{B}^{(t)}_j)\!\!\right)\!\right\}\!, 
\end{multline*}
where $%\left\{\mathbf{B}^{(t)}\right\}=
\left\{\mathbf{B}^{(t)}_1,\mathbf{B}^{(t)}_2,\cdots,\mathbf{B}^{(t)}_J\right\}$ is an arbitrary feasible fixed point.
%\end{corollary}
\end{lemma}
\begin{corollary} \label{coro-1}
A concave lower bound for $r_{k}$ is $\tilde{r}_{k}=\sum_{i}\tilde{r}_{ki}$, where $\tilde{r}_{ki}$ is a concave lower bound for $r_{ki}$ as
\begin{multline*}%{equation}
%\label{l-r-lk-p}
 r_{ki}\geq \tilde{r}_{ki}=
r_{ki_1}\left(\{\mathbf{P}\}\right) 
-r_{ki_2}^{(t-1)}
\\
-
\sum_{j=1,\neq k}^{K}
\mathfrak{R}\left\{
\text{\em Tr}\left(
\frac{
%\underline
{\mathbf{H}}_{ki}^H
(\mathbf{D}_{ki}^{(t-1)})^{-1}
%\left(\{\mathbf{P}\}, \{\bm{\Theta}^{(t-1)}\}\right) 
%\underline
{\mathbf{H}}_{ki}
}
{\ln 2}
\left(\mathbf{P}_{j}-\mathbf{P}_{j}^{(t-1)}\right)
\right)
\right\}
,
\end{multline*}
where $r_{ki_2}^{(t-1)}=r_{ki_2}\left(\{\mathbf{P}^{(t-1)}\}\right)$, %and 
$\mathbf{D}_{ki}^{(t-1)}=\mathbf{D}_{ki}\left(\{\mathbf{P}^{(t-1)}\}\right)$.
\end{corollary}
Substituting the concave lower bounds for the rates in \eqref{ar-opt-p} yields the following convex optimization problem
\begin{align}\label{ar-opt-p-sur}
 \underset{\{\mathbf{P}\}\in\mathcal{P},r
 }{\max}  & 
  r &
  \text{s.t.}  \,\,\,&  \tilde{r}_k\left(\left\{\mathbf{P}\right\}\!,\!\{\bm{\Theta}^{(t-1)}\}\right)\geq r,\,\forall k,
 \end{align}
which can be efficiently solved by existing numerical optimization tools. The solution of \eqref{ar-opt-p-sur} is the new set of the transmit covariance matrices, i.e., $\left\{\mathbf{P}^{(t)}\right\}$, which is utilized in the next step.
%\vspace{-.3cm}
\subsection{Optimizing RIS elements}%\vspace{-.2cm}
For given transmit covariance matrices, $\left\{\mathbf{P}^{(t)}\right\}$, \eqref{ar-opt} is equivalent to %\vspace{-.1cm}
%\begin{subequations}\label{ar-opt}
\begin{align}\label{ar-opt-t}
 \underset{\bm{\Theta}\in\mathcal{T},r
 }{\max}  & 
 r &
  \text{s.t.}   \,&  r_k\left(\left\{\mathbf{P}^{(t)}\right\}\!,\!\bm{\Theta}\right)\geq r,\,\forall k,%\vspace{-.2cm}
 \end{align}
%\end{subequations}
which is non-convex since $r_k\left(\left\{\mathbf{P}^{(t)}\right\}\!,\!\bm{\Theta}\right)$ is not concave in $\bm{\Theta}$, and moreover,  $\mathcal{T}$ is not a convex set. 
To solve \eqref{ar-opt-t}, we employ an MM-based approach here as well. That is, we first find suitable surrogate functions for the rates, and then convexify the set $\mathcal{T}$. %if it is not a convex set.
 To obtain a concave lower bound for the rates, we employ \cite[Lemma 2]{soleymani2022improper}, which is restated in the following lemma.
\begin{lemma}\label{lem-2} 
Consider arbitrary matrices $\mathbf{V}$ and $\bar{\mathbf{V}}$, and positive definite matrices $\mathbf{Y}$ and $\bar{\mathbf{Y}}$. Then we have:
\begin{multline} 
\ln \left|\mathbf{I}+\mathbf{V}\mathbf{V}^H\mathbf{Y}^{-1}\right|\geq
 \ln \left|\mathbf{I}+\bar{\mathbf{V}}\bar{\mathbf{V}}^H\bar{\mathbf{Y}}^{-1}\right|
\\-
\text{{\em Tr}}\left(
\bar{\mathbf{V}}\bar{\mathbf{V}}^H\bar{\mathbf{Y}}^{-1}
\right)
+
2\mathfrak{R}\left\{\text{{\em Tr}}\left(
\bar{\mathbf{V}}^H\bar{\mathbf{Y}}^{-1}\mathbf{V}
\right)\right\}\\
%&\hspace{.4cm}
-
\text{{\em Tr}}\left(
(\bar{\mathbf{Y}}^{-1}-(\bar{\mathbf{V}}\bar{\mathbf{V}}^H + \bar{\mathbf{Y}})^{-1})^H(\mathbf{V}\mathbf{V}^H+\mathbf{Y})
\right).
\label{lower-bound}
\end{multline}
\end{lemma}
\begin{corollary} \label{theo1}
A concave lower-bound for the rate of users is $\hat{r}_{k}=\sum_i\hat{r}_{ki}$, where $\hat{r}_{ki}$ is a concave lower bound for ${r}_{ki}$ as
%\begin{figure*}[t]
\begin{multline} 
%\nonumber
r_{ki}\geq
 \hat{r}_{ki}=r_{ki}^{(t-1)}-
\frac{1}{\ln 2}\left(\text{{\em Tr}}\left(
\bar{\mathbf{S}}_{ki}\bar{\mathbf{D}}_{ki}^{-1}
\right)\right.
\\
-
\text{{\em Tr}}\left(
(\bar{\mathbf{D}}^{-1}_{ki}\!-\!(\bar{\mathbf{S}}_{ki} + \bar{\mathbf{D}}_{ki})^{-1})^H (\mathbf{S}_{ki}+\mathbf{D}_{ki})
\right)
\\
\left.
+
2\mathfrak{R}\left\{
\text{{\em Tr}}\left(
\bar{\mathbf{V}}_{ki}^H\bar{\mathbf{D}}_{ki}^{-1}\mathbf{V}_{ki}
\right)\right\}
\right)
%\label{lower-bound-th1}
\end{multline}
%\hrulefill %\vspace{-.4cm}
%\end{figure*}
where $r_{ki}^{(t-1)}=r_{ki}\left(%\{\mathbf{P}^{(t)}\},
\bm{\Theta}^{(t-1)}\right)$, %and 
$%\begin{align*}
\mathbf{V}_{ki}%&
=%\underline
{\mathbf{H}}_{ki}\left(\bm{\Theta}\right)\mathbf{P}_{ki}^{(t)^{1/2}}$,
%& \hspace{1cm}
%\\
and 
$\bar{\mathbf{V}}_{ki}=%\underline
{\mathbf{H}}_{ki}\left(\bm{\Theta}^{(t-1)}\right)\mathbf{P}_{ki}^{(t)^{1/2}}$.
%\\
% \mathbf{Y}_{lk,li}&=\mathbf{D}_{lk,i}\left(\{\mathbf{P}^{(t)}\},\{\bm{\Theta}\}\right), 
%%& \hspace{1cm} 
%\\
%\bar{\mathbf{Y}}_{lk,li}&=\mathbf{D}_{lk,i}\left(\{\mathbf{P}^{(t)}\},\{\bm{\Theta}^{(t-1)}\}\right).
%\end{align*}
\end{corollary}
Substituting the concave lower bounds for the rates in \eqref{ar-opt-t} yields the following surrogate problem
%\begin{subequations}\label{ar-opt}
\begin{align}\label{ar-opt-t-sur}
 \underset{\bm{\Theta}\in\mathcal{T},r
 }{\max}  & 
  \,\,\,r &
  \text{s.t.}   \,&  \hat{r}_k\left(\left\{\mathbf{P}^{(t)}\right\}\!,\!\bm{\Theta}\right)\geq r,\,\forall k,
 \end{align}
which is still non-convex because of the unit modulus constraint in \eqref{t-3}. 
%To find a suboptimal solution for   
We can rewrite the unit modulus constraint $|\theta_m|=1$ as %the two following constraints
\begin{align}\label{14}
|\theta_m|^2&\leq 1\\
|\theta_m|^2&\geq 1.
\label{15}
\end{align}
The constraint \eqref{14} is convex. However, \eqref{15} is not a convex constraint since $|\theta_m|^2$ is a convex function (instead of being a concave function). To approximate \eqref{15} with a convex constraint, we can employ the convex-concave procedure (CCP) and relax the constraint to make the convergence faster as \cite{soleymani2022improper}
\begin{equation}\label{17}
2\mathfrak{R}\left\{
\theta_{m}^{(t-1)^*}\theta_{m}\right\}-
|\theta_{m}^{(t-1)}|^2\geq 1-\epsilon \hspace{1cm} \forall m,
\end{equation}
where $\epsilon>0$. Substituting the constraints \eqref{14} and \eqref{17} in \eqref{ar-opt-t-sur} yields the following convex problem
\begin{subequations}\label{ar-opt-18}
\begin{align}%\label{ar-opt-t-sur}
 \underset{\bm{\Theta},r
 }{\max}  & 
  r &
  \text{s.t.}   \,&  \hat{r}_k\left(\left\{\mathbf{P}^{(t)}\right\}\!,\!\bm{\Theta}\right)\geq r,\,\forall k,\\
&&&\eqref{14},\eqref{17},
 \end{align}
\end{subequations}
which can be efficiently solved. 
Let us call the solution of \eqref{ar-opt-18} as $\bm{\Theta}^\star$. 
It might happen that $\bm{\Theta}^\star$ does not meet the unit modulus constraint in \eqref{t-3} due to the relaxation in \eqref{17}. 
To generate a feasible solution, we normalize $\bm{\Theta}^\star$ as $\hat{\theta}_m=\frac{\theta^\star_m}{|\theta^\star_m|}$ for all $m$. %$\hat{\bm{\Theta}}=\frac{\bm{\Theta}^\star}{|\bm{\Theta}^\star|}$. 
Finally, we update $\bm{\Theta}$ as
\begin{equation}\label{eq-19}
%\{\bm{\Theta}^{r^{(t)}}\!\!,\bm{\Theta}^{t^{(t)}}\!\}\!\!
\bm{\Theta}^{(t)}\!\!=\!\!
\left\{\!\!\!
\begin{array}{lcl}
\hat{\bm{\Theta}}\!\!&\!\!\!\!\!\!\text{if}\!\!\!\!\!&
\min_{k}\left\{r_k\!\!\left(\hat{\bm{\Theta}}\right)\right\}\!\!\geq\!\min_{k}\left\{r_k\!\!\left(\bm{\Theta}^{(t-1)}\!\right)\right\}\!\!\!
%\\
%&&
%\min_{k}\left\{r_k\!\!\left(\bm{\Theta}^{(t-1)}\!\right)\right\}
\\
\bm{\Theta}^{(t-1)}&&\text{otherwise},
\end{array}
\right.
\end{equation}
where $\hat{\bm{\Theta}}=\text{diag}(\hat{\theta}_1,\hat{\theta}_2,\cdots,\hat{\theta}_{N_{\mathsf{RIS}}})$.
The updating policy in \eqref{eq-19} ensures convergence since the algorithm generates a non-decreasing sequence of minimum rates.

%\vspace{-.3cm}
\section{Numerical results}
In this section, we present some numerical results by employing Monte Carlo simulations. We assume that the small-scale fading for the channels ${\bf G}_i$ and ${\bf G}_{ki}$ for all $i,k$ is Rician  similar to \cite{pan2020multicell, soleymani2022rate} since there is a line of sight (LoS) link between the BS and RIS as well as between the RIS and the users. However, as  the links between the BS and the users are assumed to be non-LoS (NLoS), the small-scale fading for ${\bf F}_{ki}$ for all $k,i$ is assumed to be Rayleigh distributed. 
Note that it is also assumed that there is no correlation between the channels at different subbands. 
For more descriptions on the simulation parameters, we refer the reader to \cite{soleymani2022rate, soleymani2022noma}.

\begin{figure}[t!]
       \includegraphics[width=.42\textwidth]{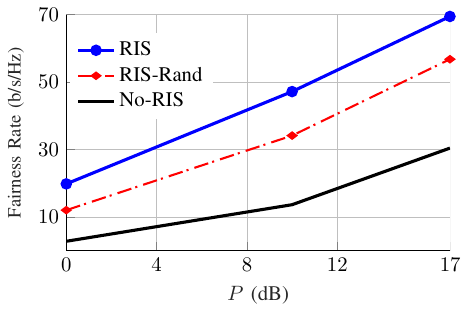}
    \caption{The average fairness rate versus $P$ for the case without RIS (No-RIS), randomly configured RIS (RIS-Rand), and optimized RIS (RIS) with $N_{\mathsf{BS}}=N_{\mathsf{U}}=2$, $K=3$, $N_{\mathsf{RIS}}=100$,  and $N_i=16$.}
	\label{Fig-rr4} 
\end{figure}
Fig. \ref{Fig-rr4} shows the average fairness rate versus $P$ for $N_{\mathsf{BS}}=N_{\mathsf{U}}=2$, $K=3$, $N_{\mathsf{RIS}}=100$, and $N_i=16$. As can be observed, RIS can significantly improve the system performance even when the RIS elements are randomly chosen.
Additionally, it can be observed that there is almost a constant gap between the proposed scheme for RIS-assisted systems and the scheme with random RIS coefficients (RIS-Rand). Indeed, even though the RIS elements cannot be independently optimized at each subband, we can get a significant gain by optimizing RIS coefficients, which shows the effectiveness of RIS in multi-user MIMO OFDM systems.

\begin{figure}[t!]
    \centering
    \begin{subfigure}[t]{0.24\textwidth}
        \centering
       \includegraphics[width=\textwidth]{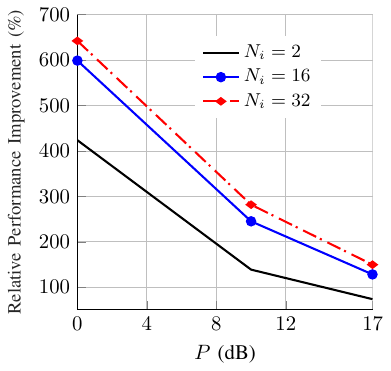}%{fig/con2}
        \caption{RIS compared to No-RIS.}
    \end{subfigure}%
    ~
    \begin{subfigure}[t]{0.24\textwidth}
        \centering
           \includegraphics[width=\textwidth]{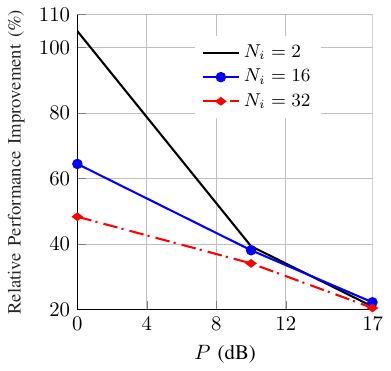}%{fig/con3}
        \caption{RIS compared to RIS-Rand.}
    \end{subfigure}
    \caption{The average improvements by RIS versus $P$ for $N_{\mathsf{BS}}=N_{\mathsf{U}}=2$, $K=3$,  $L=2$,  $M=2$, and $N_i=16$.}%\vspace{-.5cm}
	\label{Fig-rr3} 
\end{figure}
Fig. \ref{Fig-rr3} shows the average performance improvements versus $P$ for $N_{\mathsf{BS}}=N_{\mathsf{U}}=2$, $K=3$, and different $N_i$. The relative performance curves in Fig. \ref{Fig-rr3}a and Fig. \ref{Fig-rr3}b are obtained by comparing  the average fairness rate of our proposed scheme for RIS-assisted OFDM systems with the average fairness rate of OFDM systems without RIS and with the average fairness rate of OFDM systems with random RIS coefficients, respectively. As can be observed, RIS can provide a huge gain. However, the benefits of optimizing RIS components highly decrease with $N_i$, as shown in Fig. \ref{Fig-rr3}b. 
The reason is that RIS components cannot be independently optimized at each subband, and as the number of subbands for a fixed $N_{\mathsf{RIS}}$ increases, the effectiveness of optimizing $\bm{\Theta}$ decreases.  
Interestingly, the benefits of optimizing RIS elements are still significant even when there are slightly higher than $1$ RIS elements per user per subband ($N_{\mathsf{RIS}}/(KN_i)\simeq 1.04$ when $N_i$=32).  Furthermore, we observe that the benefits of optimizing RIS elements are much higher in low SNR regimes. 
Since we also consider a power/covariance matrix optimization, it may happen that the signals for a user are transmitted over a few number of subbands when the BS power budget is low. Thus, the effective number of subbands is lower than $N_i$, especially at low SNR regimes, which enhances the benefits of optimizing RIS elements.  Additionally, as shown in our previous studies \cite{soleymani2022improper,soleymani2022rate}, the benefits of employing RIS are higher in low SNR regimes, which may enhance the gain of a proper optimization of RIS elements.
%\vspace{-.4cm}
\section{Conclusion}
%\vspace{-.1cm}
In this paper, we have proposed a precoding and RIS element optimization scheme for a MIMO RIS-assisted OFDM BC to maximize the minimum rate of users. We showed that RIS can significantly improve the system performance even when the number of RIS components per subbands is very low. Moreover, our numerical results show that the benefits of optimizing RIS elements in low SNR regimes are much higher than high SNR regimes. Additionally, we showed that the benefits of optimizing RIS elements are still significant in OFDM systems even though the RIS coefficients cannot be optimized at each subband independently. %\vspace{-.1cm}
%\newpage
%\vspace{-.2cm}
\bibliographystyle{IEEEtran}
\bibliography{ref2}
\end{document}